\documentclass[prb,twocolumn,showpacs,amsmath,amssymb,superscriptaddress,preprintnumbers]{revtex4}
\usepackage{graphicx}
\usepackage{bm}

\newcommand{\ij}{i\kern -0.08em j}
\newcommand{\Oc}{\mathcal{O}}
\newcommand{\half}{{\textstyle\frac{1}{2}}}
\newcommand{\phz}{{\vphantom{z}}}
\newcommand{\ket}[1]{\lvert#1\rangle}
\newcommand{\hot}{+\Oc[(E/\hn E_0)^2]}
\newcommand{\dlf}{\delta\!f}
\def\hn{\mskip-0.5\thinmuskip}
\def\hp{\mskip0.5\thinmuskip}

\begin{document}

\title{Direct Josephson coupling between superconducting flux qubits}
\author{M.~Grajcar}
\affiliation{%
Institute for Physical High Technology, P.O. Box 100239, D-07702
Jena, Germany}
\affiliation {%
Department of Solid State Physics,
Comenius University, SK-84248 Bratislava, Slovakia}
\author{A.~Izmalkov}
\affiliation{%
Institute for Physical High Technology, P.O. Box 100239, D-07702
Jena, Germany}
\affiliation {%
Moscow Engineering Physics Institute (State University),
Kashirskoe shosse 31, 115409 Moscow, Russia}
\author{S.H.W.~van~der~Ploeg}
\affiliation{%
Institute for Physical High Technology, P.O. Box 100239, D-07702
Jena, Germany}
\affiliation{MESA+ Research Institute and Faculty of
Science and Technology, University of Twente, P.O. Box 217, 7500
AE Enschede, The Netherlands}
\author{S.~Linzen}
\author{E.~Il'ichev}
\email{ilichev@ipht-jena.de}
\affiliation{%
Institute for Physical High Technology, P.O. Box 100239, D-07702
Jena, Germany}
\author{Th.~Wagner}
\affiliation{%
Institute for Physical High Technology, P.O. Box 100239, D-07702
Jena, Germany}
\author {U.~H\"ubner}
\affiliation{%
Institute for Physical High Technology, P.O. Box 100239, D-07702
Jena, Germany}
\author{H.-G.~Meyer}
\affiliation{%
Institute for Physical High Technology, P.O. Box 100239, D-07702
Jena, Germany}
\author{Alec Maassen van den Brink}
\email{alec@dwavesys.com}
\affiliation{%
D-Wave Systems Inc., 320-1985 W. Broadway, Vancouver, V6J 4Y3 Canada}
\author{S.~Uchaikin}
\affiliation{%
D-Wave Systems Inc., 320-1985 W. Broadway, Vancouver, V6J 4Y3 Canada}
\affiliation{%
Institute for Physical High Technology, P.O. Box 100239, D-07702
Jena, Germany}
\author{A.M. Zagoskin}\thanks{Currently at UBC; email: \texttt{zagoskin@physics.ubc.ca}}
\affiliation{%
D-Wave Systems Inc., 320-1985 W. Broadway, Vancouver, V6J 4Y3
Canada} \affiliation{Physics and Astronomy Dept., The University
of British Columbia,  6224
Agricultural Rd., Vancouver, V6T 1Z1 Canada}

\date{\today}

\begin{abstract}
We have demonstrated strong antiferromagnetic coupling between two
three-junction flux qubits based on a shared Josephson
junction, and therefore not limited by the small inductances of the
qubit loops. The coupling sign and magnitude were measured by
coupling the system to a high-quality superconducting tank
circuit. Design modifications allowing to continuously tune
the coupling strength and/or make the coupling ferromagnetic are discussed.

\end{abstract}


\maketitle

Quantum superposition of macroscopic states was conclusively demonstrated in superconducting Josephson structures in 2000.\cite{cat} Such structures are natural candidates for the role of qubits (quantum bits), the constituent elements of quantum computers. Successful operation of a  quantum computer would be the ultimate confirmation of the validity of quantum mechanics on the macroscopic scale, which makes the task of controllably linking a significant number of  qubits together more than just an advance in technology.

The coupling energy~$J$ must be comparable to the splittings between the two lowest eigenstates of individual qubits. On the other hand, the coupling must not excite the qubits to higher levels, or significantly increase the qubits' interaction with undesirable degrees of freedom, leading to decoherence and dissipation. Finally, $J$ should be either variable by design or, even better, tunable during the system's operation.

In this letter we demonstrate the coupling of two three-Josephson-junction (3JJ) flux qubits, making progress towards meeting these requirements, and discuss the ways of its further improvement. The 3JJ qubit consists of a superconducting loop with small inductance~$L$ interrupted by three Josephson junctions. The two different directions of persistent current in the loop form the qubit's basis states.\cite{Mooij99} The 3JJ design enables classical bistability even for $L\rightarrow0$, resulting in a weak coupling to environmental magnetic flux noises. As a result, quantum behaviour with long decoherence time was observed in this type of qubit by several groups.\cite{Chiorescu03,Ilichev03,Bertet04}

\setlength{\unitlength}{1mm}
\begin{figure}[tbp]
\begin{picture}(70,30)
  \put(7,0){\line(1,0){56}}
  \put(7,30){\line(1,0){56}}
  \put(7,0){\line(0,1){30}}
  \put(63,0){\line(0,1){30}}
  \put(35,0){\line(0,1){30}}
  \put(31,11){\line(1,1){8}}
  \put(31,19){\line(1,-1){8}}
  \put(4,2){\line(1,1){6}}
  \put(4,8){\line(1,-1){6}}
  \put(4,22){\line(1,1){6}}
  \put(4,28){\line(1,-1){6}}
  \put(60,2){\line(1,1){6}}
  \put(60,8){\line(1,-1){6}}
  \put(60,22){\line(1,1){6}}
  \put(60,28){\line(1,-1){6}}
  \put(5,13){\line(1,1){4}}
  \put(5,17){\line(1,-1){4}}
  \put(61,13){\line(1,1){4}}
  \put(61,17){\line(1,-1){4}}
  \put(20,13.8){$f_a$}
  \put(48,13.8){$f_b$}
  \put(30,13.8){0}
  \put(3,23.8){1}
  \put(3,13.8){2}
  \put(3,3.8){3}
  \put(65,23.8){4}
  \put(65,13.8){5}
  \put(65,3.8){6}
\end{picture}
\caption{Schematics of a 7JJ device: two 3JJ qubits with direct Josephson coupling and reduced flux bias~$f_{a;b}$.} \label{figA}
\end{figure}
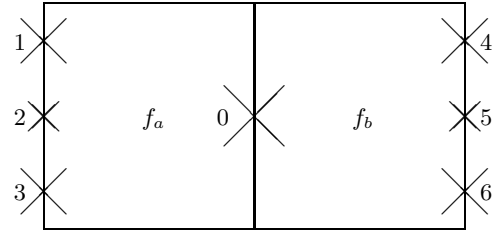

However, their small $L$ makes it difficult to couple 3JJ qubits inductively; generally, $J$ is smaller than the single-qubit level splitting. We therefore implement the proposals\cite{Levitov01,Butcher02} to directly link two qubits through a shared junction (Fig.~\ref{figA}). The resulting coupling not only is strong, but can also be varied independently of other design parameters by choosing the shared junction's size.

To calculate~$J$, we neglect the inductances so that the potential term in the Hamiltonian contains only the Josephson energy $U_\mathrm{J} = -\sum_{j=0}^6 E_j \cos\phi_j$, and use flux quantization,
$\phi_1+\phi_2+\phi_3+\phi_0 = 2\pi(\half{+}f_a)$ and
$\phi_4+\phi_5+\phi_6-\phi_0 = 2\pi(\half{+}f_b)$, to eliminate~$\phi_{2;5}$ ($f_{a;b}=\Phi^\mathrm{x}_{a;b}/\Phi_0-\half$ is the reduced flux bias). In the simplest case [$f_{a;b} = 0$; $E_{1;3;4;6} = E$; $E_{2;5} = \alpha E$ ($\half < \alpha < 1$); $E_0 \gg E$] there are two different pairs of potential minima: ``ferro-'' and ``antiferromagnetic'' (with parallel and antiparallel loop currents respectively),
\begin{align}
  \phi_0^\mathrm{FM} &= 0\;, \notag\\
  \phi_{1;3;4;6}^\mathrm{FM} &= \pm\arccos(1/2\alpha)\;;\label{eq_FM}\\
  \phi_0^\mathrm{AF} &= \pm \frac{\hbar I_\mathrm{p}}{eE_0}\hot\;,\notag\\
  \phi_{1;3}^\mathrm{AF} = -\phi_{4;6}^\mathrm{AF} &=
  \pm\arccos(1/2\alpha)\pm \frac{1{-}2\alpha^2}{4\alpha^2{-}1}
   \frac{\hbar I_\mathrm{p}}{eE_0}\notag\\
   &\quad\,\hot\;,\label{eq_AF}
\end{align}
where $I_\mathrm{p} = (2e/\hbar)E\sqrt{1-1/4\alpha^2}$ is the persistent current in a free 3JJ qubit.\cite{Mooij99} Inserting these into~$U_\mathrm{J}$, one finds that the AF states have the lower energy by
\begin{equation}
  \Delta U = 2J = \frac{\hbar^2I_\mathrm{p}^2}{2e^2E_0}\hot\;,\label{eq_DeltaU1}
\end{equation}
so that the effective mutual inductance $\hbar^2\!/4e^2E_0$ is just the standard Josephson inductance of the coupling junction.\cite{general} For an explanation, note that flipping the signs of $\phi_{4;6}^\mathrm{FM}$ yields an AF configuration with $\phi_0=0$, and with the same energy as the FM minimum. Such a state of course is non-stationary (since charge must be conserved), and adjustment of the phases will lower~$U_\mathrm{J}$, with the stationary AF state (\ref{eq_AF}) realizing the global minimum. More intuitively: the nonzero $\phi_0^\mathrm{AF}$ reduces the effective frustration in the individual qubits, which is maximal for $f_{a;b} = 0$ [cf.\ above~(\ref{eq_FM})]; the attendant reduction in qubit energy overcomes (by a factor two) the increase in Josephson energy in the coupling junction itself.

Thus the direct Josephson coupling of 3JJ qubits has the same sign as their inductive coupling (the latter corresponding to the natural north--south alignment of their magnetic moments),\cite{JJ-ind} but is not restricted by the geometric inductances. For $E_0\to\infty$, $J$ in (\ref{eq_DeltaU1}) disappears as it should, since then we have two qubits sharing a common leg without a junction---equivalent to two adjacent qubits if kinetic inductance is neglected, with only inductive coupling.~\cite{majer}

\begin{figure}[tbp]
  \includegraphics[width=8cm]{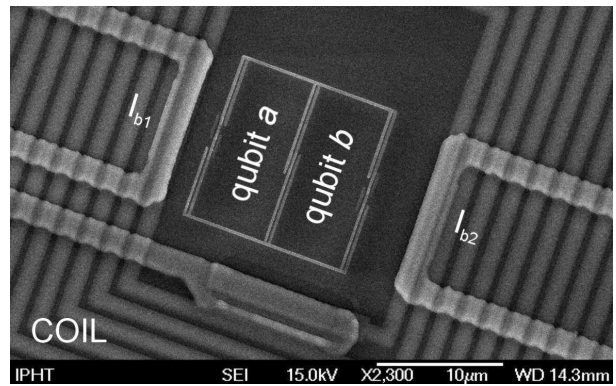}
\caption{Micrograph of sample~1 (see the table). The 3JJ qubits $a$ and $b$ are coupled through a shared Josephson junction, visible in the centre, and can be flux-biased independently by dc lines $I_\mathrm{b1}$ and $I_\mathrm{b2}$. The pancake coil is part of an $LC$ circuit reading out the qubits' magnetic susceptibility.} \label{Levels}
\end{figure}

We determined $J$ using an impedance measurement technique, applied previously to 3JJ qubits\cite{Ilichev03,Grajcar04} and extended to multiple qubits in Ref.~\onlinecite{Izmalkov04}. The qubits are placed inside a tank circuit with known self-inductance~$L_\mathrm{T}$ and quality factor~$Q_\mathrm{T}$, driven by a dc bias plus a small ac current at its resonance frequency~$\omega_\mathrm{T}$ (Fig.~\ref{Levels}). The tank's voltage--current phase angle~$\theta$ is given by $\tan\theta=-(Q_\mathrm{T}/L_\mathrm{T})\chi'$. Here, $\chi'$ is the qubits' contribution to the tank susceptibility,\cite{chi} readily related to the curvature of their energy bands: for qubit--tank mutual inductances $M_{a\mathrm{T}}=M_{b\mathrm{T}}\equiv M$, one has\cite{greenberg}
\begin{equation}
  \tan\theta = \frac{Q_\mathrm{T}}{L_\mathrm{T}}M^2
  \frac{d^2E_\mathrm{tot}}{d(\Phi^\mathrm{x})^2}\;, \label{Eq:theta}
\end{equation}
where $\Phi^\mathrm{x}$ denotes a symmetric change of flux bias in both qubit loops. At temperature $T=0$, $E_\mathrm{tot}$ is the qubits' ground-state energy; at finite~$T$, the derivative simply becomes a Boltzmann average over the levels. The band curvature is large near anticrossings, so that $\tan\theta(f_a,f_b)$ contains important information about the level structure.

One obtains $J$ from such a plot as follows. The standard four-state Hamiltonian for two coupled qubits is
\begin{equation}\label{eq_Hamiltonian}
  H=-\epsilon_a^\phz \sigma_a^z-\Delta_a^\phz \sigma_a^x
   -\epsilon_b^\phz\sigma_b^z-\Delta_b^\phz \sigma_b^x+J\sigma_a^z\sigma_b^z\;,
\end{equation}
where $\sigma^x$, $\sigma^z$ are Pauli matrices, $\Delta_j$ is the tunneling amplitude, and $\epsilon_j=I_{\mathrm{p}j}\Phi_0f_j$ is the energy bias ($j=a,b$). For low $T$ and small~$\Delta_j$, the \emph{location} of the peaks in $\lvert\tan\theta\rvert$ (due to anticrossings) follows simply from the classical stability diagram. For instance, the $\ket{\uparrow\downarrow}\leftrightarrow\ket{\uparrow\uparrow}$ transition ($\ket{\uparrow\downarrow}$: $\sigma_a^z=-\sigma_b^z=1$ etc.) occurs at $-\epsilon_a+\epsilon_b-J=-\epsilon_a-\epsilon_b+J\Rightarrow\epsilon_b=J$. Therefore, the peak-to-peak distance equals~$2J$.

For our samples, we first fabricate niobium (Nb) pancake coils and dc flux-bias lines on 4-inch oxidized silicon wafers, and then the qubits inside the coils by aluminium (Al) shadow evaporation on 12$\times$12 mm$^{2}$ chips.

The Nb process starts with sputtering and dry etching of the 200~nm thick coil windings with 1~$\mu$m width, 1~$\mu$m line spacing, and typically 30 turns. The patterning uses $e$-beam lithography and a CF$_4$ RIE process. Then a silicon oxide isolation layer and the second 300~nm thick Nb film are deposited for the central coil electrode and the 2~$\mu$m wide dc lines; photolithography is used for all required resist masks of these layers. Finally, 400~nm silicon oxide is deposited for protection and isolation.

The Al process uses $e$-beam lithography to prepare the double layer resist mask for the qubits with a 150~nm linewidth. The two Al layers are deposited in situ by $e$-beam evaporation with different angles of incidence at a rate of 1.8~nm/s. The surface of the first Al film is oxidized with pure oxygen at a pressure of $10^{-2}$~mbar. The qubits are completed after the final lift-off.

\begin{figure}[tbp]
  \includegraphics[width=8cm]{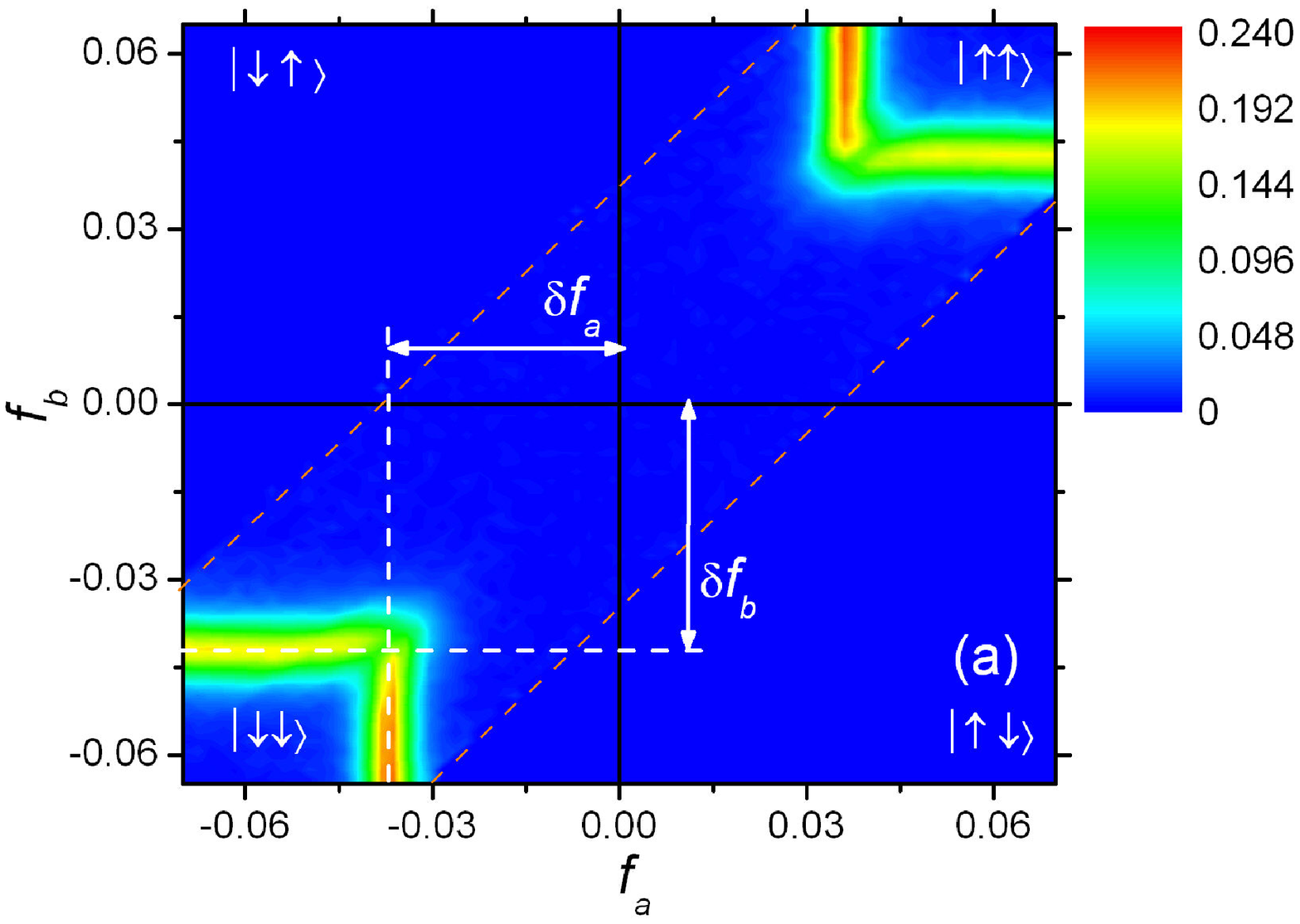}
  
  \vspace{2mm}
  \includegraphics[width=8cm]{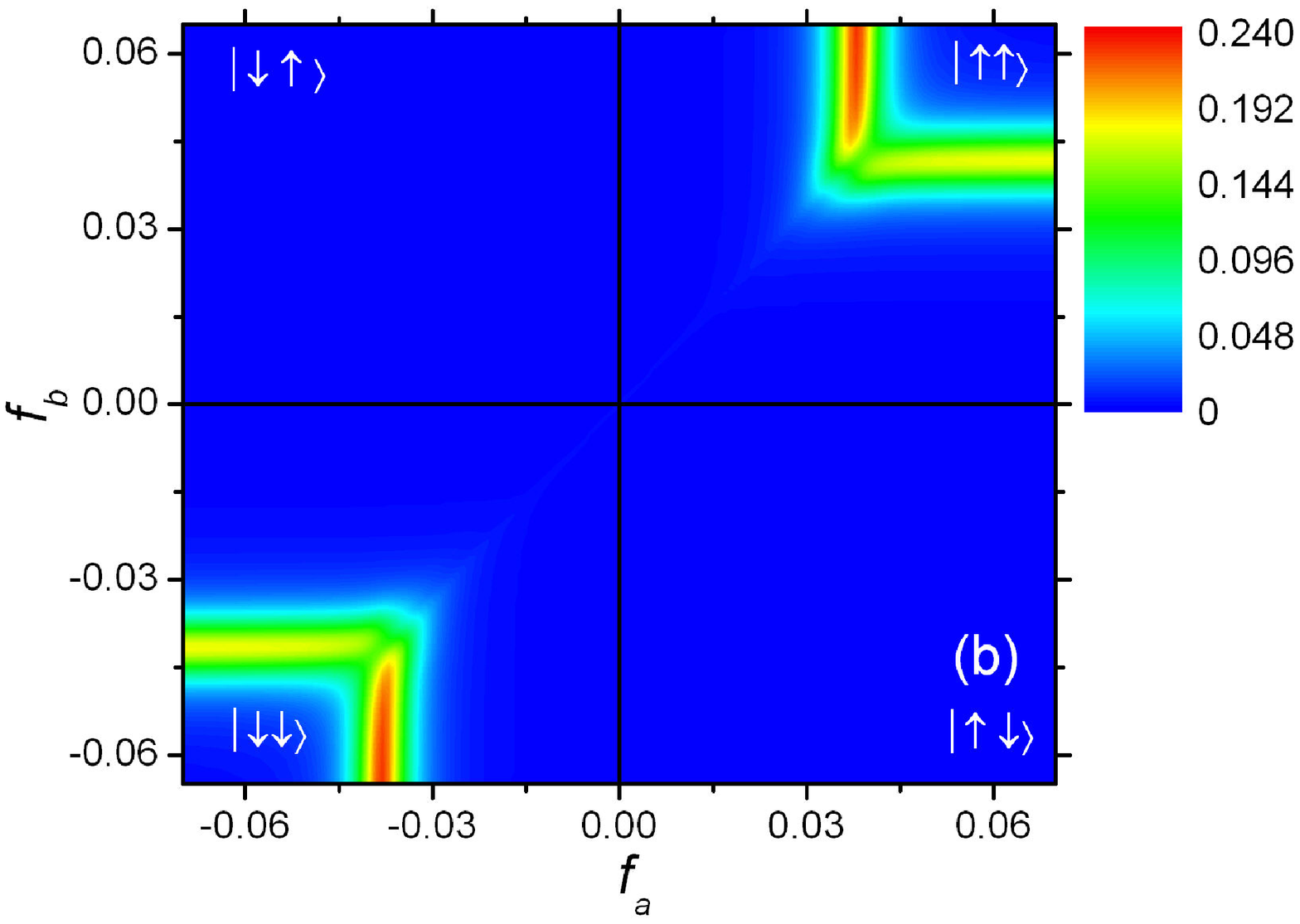}
\caption{Plot of $-\tan\theta(f_a,f_b)$ for two coupled 3JJ qubits
(sample 1 from the table). (a): Measurements at a nominal mixing-chamber temperature of 20~mK (actual data taken between the dashed lines). (b): Theoretical fit for $T=70$~mK (see the text for the different $T$'s). The coupling strength can be estimated from the separation between the peaks.}
\label{Disney}
\end{figure}

Results are shown in Fig.~\ref{Disney}(a). As explained below~(\ref{eq_Hamiltonian}), one has $J\approx I_{\mathrm{p}a}\Phi_0\hp\dlf_a=I_{\mathrm{p}b}\Phi_0\hp\dlf_b$. One can find $I_{\mathrm{p}a;b}$ in two different ways, which agree to within~$\sim$20\%. First, we used $I_\mathrm{p}\approx I_\mathrm{c}\sqrt{1-1/4\alpha^2}$ [cf.\ below (\ref{eq_AF})]. Here, $I_\mathrm{c}$ is the critical current of a junction fabricated on the same chip and with the same area of 650$\times$150~nm$^2$ as junctions 1/3/4/6, enclosed in a superconducting loop and measured by the conventional rf-SQUID technique; $\alpha\approx0.75$ by design. The second way is to fit the \emph{shape} of the peaks in $\tan\theta(f_a,f_b)$ [Fig.~\ref{Disney}(b)], using the spectrum of (\ref{eq_Hamiltonian}) to evaluate (\ref{Eq:theta}), which yields $\Delta_j$ and~$I_{\mathrm{p}j}$.\cite{Grajcar04,Smirnov03,Izmalkov04} The required tank parameters were extracted from its resonance characteristic, and the mutual inductances from the flux periodicity; for sample~1, $L_\mathrm{T}=136$~nH, $Q_\mathrm{T}=664$, $\omega_\mathrm{T}/2\pi=19.925$~MHz, and $M_{a\mathrm{T}}=M_{b\mathrm{T}}=66.5$~pH. The $\ket{\uparrow\downarrow}\leftrightarrow\ket{\downarrow\uparrow}$ anticrossing does not show up in the figure because there is no net flux change, hence no contribution to the qubit susceptibility; one can also say that the level curvature peaks perpendicular to the symmetric direction stipulated in~(\ref{Eq:theta}).

\begin{table}
\caption{Coupling-junction areas~$S_0$, tunneling amplitudes~$\Delta_j$, persistent currents~$I_{\mathrm{p}j}$, coupling energies~$J$, and peak locations $\dlf$ for the measured samples.}\label{tab:qubitParam}

\vspace{1mm}
  \begin{tabular}{l|ccccclcc}
  sample & $S_0$  & $\Delta_a$ & $\Delta_b$ & $I_{\mathrm{p}a}$ & $I_{\mathrm{p}b}$ & $J$ & $\dlf_\mathrm{exp}$ & $\dlf_\mathrm{th}$ \\
  No.    &$\mu$m$^2$ &  mK       & mK       & nA       &  nA  & K & & \\ \hline
  1      &  0.30  &  80  &  90  &  120  &  110  &  0.7 & 0.037--0.041 & 0.0360 \\
  2      &  0.15  &  30  &  30  &  150  &  120  &  1.2  & 0.068 & 0.0675 \\
\end{tabular}
\end{table}

The results of the fit are summarized in the table for two measured two-qubit samples, with different sizes of junction~0. Note how, say, $\Delta_a<\Delta_b$ for sample~1 makes the $a$-anticrossing sharper, resulting in a deeper colour for the corresponding peak (vertical bands in Fig.~\ref{Disney}). Since $E_0/\hn E=3.1$ and 1.5, respectively, the perturbative analysis leading to (\ref{eq_DeltaU1}) does not apply quantitatively (the latter would have required large coupling junctions, which proved difficult to make with sufficient homogeneity). Instead, a theoretical prediction is made for $\dlf(E_0/\hn E,\alpha)$, by calculating the classical stability diagram directly from $U_\mathrm{J}(\phi_0,\phi_1,\phi_3,\phi_4,\phi_6)$. Incidentally, this has the advantage that the critical-current density drops out of the comparison (last two columns in the table), which therefore shows greater accuracy than we can claim for $J$ itself.

\begin{figure}[tbp]
  \includegraphics[width=8cm]{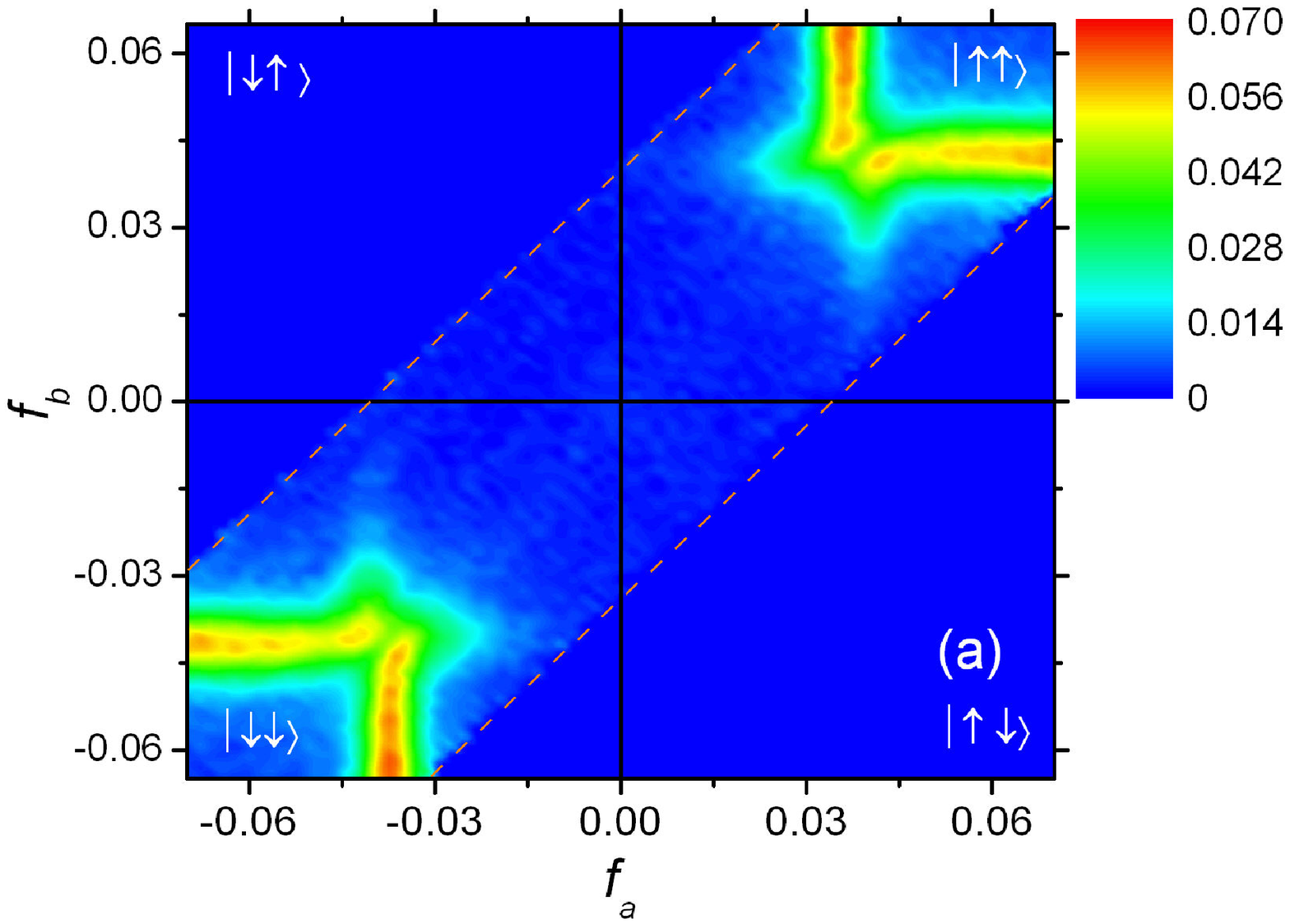}
  
  \vspace{2mm}
  \includegraphics[width=8cm]{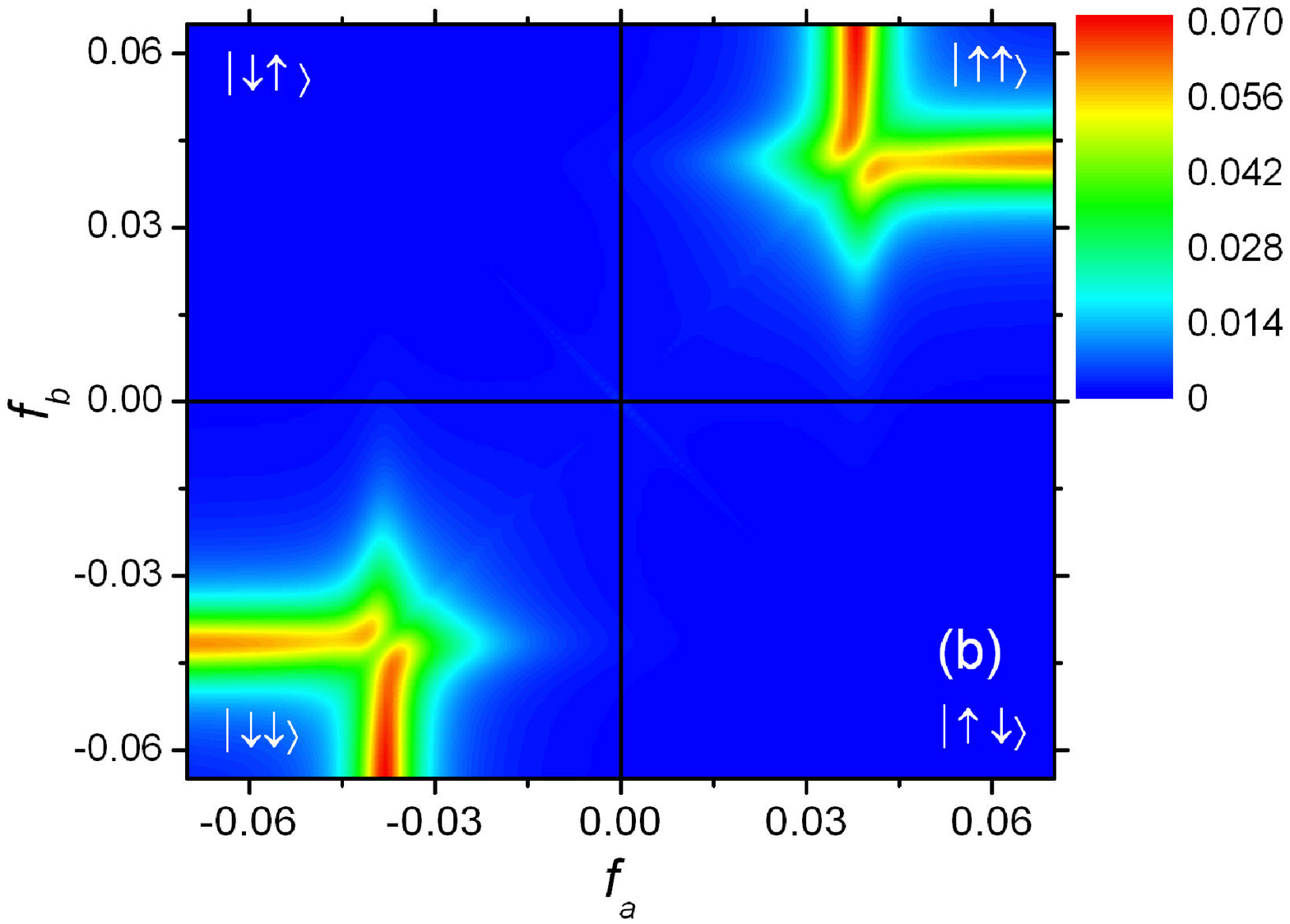}
\caption{As in Fig.~\ref{Disney}, but for a mixing-chamber temperature of 200~mK~(a) and an effective $T=300$~mK~(b).}
\label{hi-T}
\end{figure}

Data taken at a higher~$T$ support the effective Hamiltonian (\ref{eq_Hamiltonian}) for our 7JJ system beyond the ground state. Namely, for, say, $f_a\lesssim0.035$, $f_b\approx0.042$, an $\ket{\uparrow\downarrow}\leftrightarrow\ket{\uparrow\uparrow}$ anticrossing persists between excited states of sample~1. At finite~$T$, it should contribute in (\ref{Eq:theta}), with rapidly decreasing Boltzmann weight as $f_a$ is reduced. This is precisely what is seen in Fig.~\ref{hi-T}(a); the fit in Fig.~\ref{hi-T}(b) shows detailed agreement with the theory. In both Figs.~\ref{Disney} and~\ref{hi-T}, the discrepancy between effective and mixing-chamber temperatures is well within the range expected due to heating through external leads etc.; we observed no significant deviations from an equilibrium distribution.

The remarkable $J\sim1$~K significantly exceeds both the tunnel splitting and the inductive coupling (estimated to be $\sim$20~mK). It can be flux-tuned by using a standard compound junction (dc-SQUID) for the coupling. Instead, one can also apply a bias current $I_\mathrm{b}$ through junction~0.\cite{Lantz04} The corresponding generalization of (\ref{eq_DeltaU1}) reads $J = \hbar I_\mathrm{p}^2/2e\sqrt{I_\mathrm{c0}^2{-}I_\mathrm{b}^2}\hot$. Thus, $J$ can only be \emph{increased}, albeit significantly. Hence, this mechanism does not allow, e.g., changing the coupling sign and tunable decoupling of qubits. These are desirable for most quantum algorithms, but existing proposals for flux qubits rely on, and therefore are limited by, mutual inductances.\cite{Plourde04} A bias line will introduce some noise. For reference, we give the coupling linewidth due to low-frequency fluctuations in $I_\mathrm{b}$ with spectral density~$S_\mathrm{b}(\omega)$: $\Delta J=\hbar I_\mathrm{p}^4 I_\mathrm{b}^2S_\mathrm{b}(0)/4e^2(I_\mathrm{c0}^2{-}I_\mathrm{b}^2)^3$.

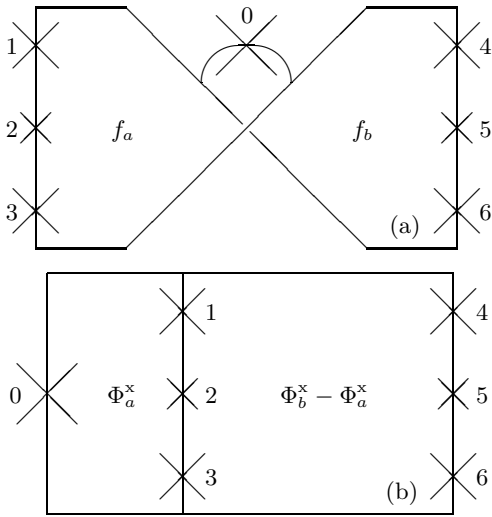
\begin{figure}[ht]
\begin{picture}(64,32)
  \put(4,0){\line(1,0){12}}
  \put(4,32){\line(1,0){12}}
  \put(48,0){\line(1,0){12}}
  \put(48,32){\line(1,0){12}}
  \put(4,0){\line(0,1){32}}
  \put(60,0){\line(0,1){32}}
  \put(16,0){\line(1,1){32}}
  \put(16,32){\line(1,-1){15.5}}
  \put(48,0){\line(-1,1){15.5}}
  \put(28,23){\line(1,1){8}}
  \put(28,31){\line(1,-1){8}}
  \put(32,22){\oval(12,10)[t]}
  \put(1,2){\line(1,1){6}}
  \put(1,8){\line(1,-1){6}}
  \put(1,24){\line(1,1){6}}
  \put(1,30){\line(1,-1){6}}
  \put(57,2){\line(1,1){6}}
  \put(57,8){\line(1,-1){6}}
  \put(57,24){\line(1,1){6}}
  \put(57,30){\line(1,-1){6}}
  \put(2,14){\line(1,1){4}}
  \put(2,18){\line(1,-1){4}}
  \put(58,14){\line(1,1){4}}
  \put(58,18){\line(1,-1){4}}
  \put(14,14.8){$f_a$}
  \put(46,14.8){$f_b$}
  \put(0,14.8){2}
  \put(0,25.8){1}
  \put(31.3,30){0}
  \put(0,3.8){3}
  \put(63,25.8){4}
  \put(63,14.8){5}
  \put(63,3.8){6}
  \put(51,2){(a)}
\end{picture}
  
\vspace{3mm}
\begin{picture}(63,32)
  \put(5,0){\line(1,0){54}}
  \put(5,32){\line(1,0){54}}
  \put(5,0){\line(0,1){32}}
  \put(59,0){\line(0,1){32}}
  \put(23,0){\line(0,1){32}}
  \put(1,12){\line(1,1){8}}
  \put(1,20){\line(1,-1){8}}
  \put(20,2){\line(1,1){6}}
  \put(20,8){\line(1,-1){6}}
  \put(20,24){\line(1,1){6}}
  \put(20,30){\line(1,-1){6}}
  \put(56,2){\line(1,1){6}}
  \put(56,8){\line(1,-1){6}}
  \put(56,24){\line(1,1){6}}
  \put(56,30){\line(1,-1){6}}
  \put(21,14){\line(1,1){4}}
  \put(21,18){\line(1,-1){4}}
  \put(57,14){\line(1,1){4}}
  \put(57,18){\line(1,-1){4}}
  \put(13,14.8){$\Phi^\mathrm{x}_a$}
  \put(36,14.8){$\Phi^\mathrm{x}_b-\Phi^\mathrm{x}_a$}
  \put(26,14.8){2}
  \put(26,25.8){1}
  \put(0,14.8){0}
  \put(26,3.8){3}
  \put(62,25.8){4}
  \put(62,14.8){5}
  \put(62,3.8){6}
  \put(50,2){(b)}
\end{picture}
\caption{Two 3JJ qubits with ferromagnetic Josephson coupling. (a):~Twisted design. (b):~Overlapping design.}
\label{FM-fig}
\end{figure}

Other variations are presented in Fig.~\ref{FM-fig}. In Fig.~\ref{FM-fig}(a), the relative twist between the qubit loops interchanges the role of the AF and FM configurations, so that the latter have the lowest energy. In particular, the strength of the direct FM Josephson coupling can overcome any residual AF inductive interaction. Junction~0 can presumably be fabricated between the crossing lines in the centre. In Fig.~\ref{FM-fig}(b), one qubit loop is 01230 and the other is 04560; by choosing 1:2 area ratios, both qubits can be brought close to degeneracy with a homogeneous field, for $\Phi^\mathrm{x}_a=\frac{1}{3}\Phi^\mathrm{x}_b=\half\Phi_0$. One obtains FM coupling without a twisted layout, but with strongly asymmetric qubits. The two discussed modifications can be combined: by current-biasing the junctions~0 in Fig.~\ref{FM-fig}, one obtains tunable FM coupling.

In conclusion, we have demonstrated direct antiferromagnetic Josephson coupling between two individually controllable three-junction flux qubits. The coupling strength can be on the order of a Kelvin. We also proposed design modifications allowing tunable and/or ferromagnetic coupling. Future experimental work should also consider linear qubit arrays, to which the design of Fig.~\ref{figA} is readily generalized.\cite{Butcher02}

EI thanks the EU for support through the RSFQubit project. SvdP thanks the ESF pi-shift programme for a grant. MG acknowledges support by Grant No.\ VEGA 1/2011/05. AMB and  AZ thank M.H.S. Amin, A.Yu.\ Smirnov, and M.F.H. Steininger for fruitful discussions.


\begin{thebibliography}{99}
\bibitem{general} By considering ``black box'' devices in the outer arms, $U_\mathrm{J}=U_a\bm{(}2\pi(\half{+}f_a){-}\phi_0\bm{)}+U_b\bm{(}2\pi(\half{+}f_b){+}\phi_0\bm{)}-E_0\cos\phi_0$, one readily verifies that the latter result also holds if the two qubits are asymmetric, biased, or of a different type.
\bibitem{chi} Properly, $\chi'$ is the real part, effective at~$\omega_\mathrm{T}$.
\end{thebibliography}
\end{document}